\documentclass[12pt,a4paper,reqno]{amsart}
\usepackage{amsmath, amsfonts,
   amssymb, 
   amsthm, euscript}

\newtheorem{theor}{Theorem}
\theoremstyle{definition}

\newtheorem{state}[theor]{Proposition}
\newtheorem{lemma}[theor]{Lemma}
\newtheorem{cor}[theor]{Corollary}

\newtheorem*{defNo}{Definition}

\newtheorem{example}{Example}

\theoremstyle{remark}
\newtheorem{rem}{Remark}

\DeclareFontFamily{OML}{cyr}{}
\DeclareFontShape{OML}{cyr}{m}{n}{
   <5> <6> <7> <8> <9> gen * wncyr
   <10> <10.95> <12> <14.4> <17.28> <20.74> <24.88> wncyr10
  }{}
\DeclareSymbolFont{rusletters}{OML}{cyr}{m}{n}
\DeclareSymbolFontAlphabet{\rusmath}{rusletters}
\DeclareMathSymbol\re{\rusmath}{rusletters}{"03}
\newcommand{\cEv}{\EuScript{E}}

\newcommand{\pinner}{\mathbin{\mathchoice
   {\hbox{\vrule width0.6em depth0pt height0.4pt
   \vrule width0.4pt depth0pt height0.8ex}}
   {\hbox{\vrule width0.6em depth0pt height0.4pt
   \vrule width0.4pt depth0pt height0.8ex}}
   {\hbox{\kern0.14em
   \vrule width0.48em depth0pt height0.4pt
   \vrule width0.4pt depth0pt height0.6ex\kern0.14em}}
   {\hbox{\kern0.1em
   \vrule width0.39em depth0pt height0.4pt
   \vrule width0.4pt depth0pt height0.5ex\kern0.1em}}}}

\newcommand{\cE}{\mathcal{E}}

\newcommand{\cP}{\mathcal{P}}

\newcommand{\gothg}{\mathfrak{g}}

\newcommand{\vph}{\varphi}
\newcommand{\dd}{\partial}
\newcommand{\Id}{{\mathrm d}}

\newcommand{\rme}{{\mathrm{e}}}
\newcommand{\rmN}{{\mathrm{N}}}

\DeclareMathOperator{\img}{im}
\DeclareMathOperator{\dom}{dom}

\DeclareMathOperator{\sym}{sym}
\DeclareMathOperator{\cosym}{cosym}
\DeclareMathOperator{\Cosym}{(co)sym}

\DeclareMathOperator{\divergence}{div}

\newcommand{\by}[1]{\textit{{#1}}}
\newcommand{\jour}[1]{\textit{{#1}}}
\newcommand{\vol}[1]{\textbf{{#1}}}
\newcommand{\book}[1]{\textrm{{#1}}}

\newcommand{\ib}[3]{ \{\!\{ {#1},{#2} \}\!\}_{{#3}} }

\title[YB($0$) and two $R$-\/brackets for Boussinesq\/-\/type hierarchy]%
{A family of second Lie algebra structures for symmetries of 
dispersionless Boussinesq 
system}

\date{March 6, 2009} 

\author[A. V. Kiselev]{Arthemy V. Kiselev} 
\thanks{
        \textit{Address}:
Mathematical Institute, University of Utrecht, P.O.Box~80.010, 3508~TA Utrecht, The Netherlands.
\textit{E-mails}: [\texttt{A.V.Kiselev},
\texttt{J.W.vandeLeur}]\texttt{\symbol{"40}uu.nl}%
}

\author[J. W. van de Leur]{Johan W. van de Leur}

\subjclass[2000]{
17B62, 
37K10, 
37K30. 
}

\keywords{Symmetries, brackets, Hamiltonian hierarchies, recursion operators,
integrable systems}

\begin{document}
\begin{abstract}
For the $3$-\/component dispersionless Boussinesq\/-\/type system, 
we construct two compatible nontrivial finite deformations for the Lie algebra structure in the sym\-met\-ry algebra.
\end{abstract}
\maketitle

\subsection*{Introduction}
In this short note 
we construct a two\/-\/parametric family of nontrivial finite deformations for the Lie bracket in the algebra of symmetries for the $3$\/-\/component dispersionless Boussinesq system of hydrodynamic type~\cite{Nutku,JK3Bous}
\begin{equation}\label{d-B}
\cE=\bigl\{u_t=ww_x+v_x,\quad v_t=-uw_x-3u_xw,\quad w_t=u_x\bigr\}.
\end{equation}
First, we establish a nontrivial property of a previously known~\cite{JK3Bous} self\/-\/adjoint 
Noether operator $A_0\colon\cosym\cE\to\sym\cE$: its image is
closed w.r.t.\ the commutation. Hence this operator, and the bi\/-\/Hamiltonian
pair $\smash{\bigl(\hat{A}_1,\hat{A}_2\bigr)}$ for~\eqref{d-B}, see~\cite{Nutku}, transfer the standard bracket~$[\,,\,]$ in $\sym\cE$ to the Lie algebra structures on their domain. We prove that the three new brackets are compatible.

The Noether operator~$A_0$ is invertible on an open dense subset of~$\cE$.
This yields two recursion operators $R_i=\smash{\hat{A}_i\circ A_0^{-1}}
\colon\sym\cE
\to\sym\cE$. The images of~$R_i$ are again closed w.r.t.\ the commutation, 
and this property is retained by their arbitrary linear combinations.
Using the `chain rule' formula~\eqref{ChainRuleFormula} for the 
bi\/-\/differential brackets on domains of the operators~$\smash{\hat{A}_i}$ 
and~$R_i$, 
we calculate the second Lie algebra structures $[\,,\,]_{R_i}$ on~$\sym\cE$.

All notions and constructions are standard~\cite{Olver,Opava}.
We stress that the concept of li\-ne\-ar compatible differential operators 
with involutive images, which we develop here, can be applied to the study of
other integrable systems with or without dispersion 
(e.g., see~\cite{TMPhGallipoli}).

\begin{rem}
Let $M$~be a smooth finite\/-\/dimensional orientable real manifold.
The construction of \emph{trivial infinitesimal} deformations
\[
[x,y]_{\rmN}\mathrel{{:}{=}}
\frac{\Id}{\Id\lambda}\Bigr|_{\lambda=0}\rme^{-\lambda\rmN}\,\bigl[\rme^{\lambda\rmN}(x),\rme^{\lambda\rmN}(y)\bigr]=
[\rmN x,y]+[x,\rmN y]-\rmN\bigl([x,y]\bigr)
\]
of the standard Lie algebra structure~$[\,,\,]$ on the tangent bundle
$TM\ni x,y$ is well\/-\/developed in the literature~\cite{YKSMagri}:
if the Nijenhuis torsion
$
[\rmN x,\rmN y]-\rmN\bigl([x,y]_{\rmN}\bigr)
$ 
for an endomorphism~$\rmN\colon\Gamma(TM)\to\Gamma(TM)$
vanishes, then the Lie brackets $[\,,\,]_{\rmN^k}$
obtained 
by iterations $\rmN^k$ of the Nijenhuis recursion~$\rmN$ are 
compatible (their linear combinations are Lie algebra structures as well).
Suppose further that $M$ is equipped with a Poisson bi\/-\/vector
$\cP\in\Gamma\bigl(\bigwedge^2(TM)\bigr)$. 
If the Nijenhuis and Poisson structures~$(\rmN,\cP)$ satisfy
two compatibility conditions~\cite{YKSMagri}, then they generate
infinite hierarchies of 
compatible Poisson structures
$\rmN^k\circ\cP$, $k\geq0$.
%
The concept of Poisson\/--\/Nijenhuis structures
admits a straightforward generalization~\cite{JKGolovko2008}
for the infinite jet bundles over smooth manifolds 
and for infinite\/-\/dimensional integrable systems of~PDE.
\end{rem}

On the contrary, in this paper we construct two \emph{nontrivial finite} deformations $[\,,\,]_{R_i}$ of the standard Lie bracket $[\,,\,]$ on~the symmetry algebra $\sym\cE$ for~\eqref{d-B}. We shall use two local recursion operators~$R_i$, $i=1,2$, whose images are closed w.r.t.\ the commutation. Thence we obtain the new bracket~$[\,,\,]_{R_i}$ through
\begin{equation}\label{YB}
\bigl[R_i\vph_1,R_i\vph_2\bigr]=R_i\bigl([\vph_1,\vph_2]_{R_i}\bigr)\qquad\text{for any $\vph_1,\vph_2\in\sym\cE$.}
\end{equation}
The construction constits of two steps.

\subsection*{Involutive distributions and linear compatibility}
First, let a linear operator $\square$ in total derivatives be either a recursion $\sym\cE\to\sym\cE$ for an evolutionary system~$\cE$, or a Noether operator $\cosym\cE\to\sym\cE$ whose arguments are the variational covectors\footnote{We label
the evolution equations upon $u^1,\ldots,u^m$ in the system~$\cE$ with the same variables $u^i$ that occur in the left\/-\/hand sides. With such convention, the cosymmetries $\psi$ equal the variational derivatives $\delta\rho/\delta u$ of the conserved densities~$\rho$. This yields the
transformation law for Noether operators under reparametrizations of $u^i$ that preserve the evolutionary form of~$\cE$.}
for~$\cE=\{F=0\}$ (`cosymmetries' $\psi=\nabla^*(1)$ such that 
$\divergence\eta=\langle1,\nabla(F)\rangle=\langle\psi,F\rangle$ 
for a conserved current $\eta=\rho\,\Id x+\cdots$ and the adjoint~$\nabla^*$ of
the arising operator~$\nabla$, see~\cite{Olver,Opava}). 
For example, all Hamiltonian operators for~$\cE$ are Noether.

Suppose further that the image of~$\square$ is closed w.r.t.\ the commutation in~$\sym\cE$: $[\img\square,\img\square]\subseteq\img\square$. The Lie algebra structure $[\,,\,]\bigr|_{\mathrm{im}\,A}$ is transferred by~$\square$ onto
the quotient $\Omega=\dom\square/\ker\square$: 
\[
\bigl[\square(\phi'),\square(\phi'')\bigr]=
  \square\bigl([\phi',\phi'']_\square),\qquad
\phi',\phi''\in\Omega.
\]
By the Leibnitz rule, two pairs of summands appear in the bracket of 
the evolutionary vector fields 
fields $\cEv_{\square(\phi')}$ and $\cEv_{\square(\phi'')}$: 
\[\bigl[\square(\phi'),\square(\phi'')\bigr]=
  \square\bigl(\cEv_{\square(\phi')}(\phi'')-
  \cEv_{\square(\phi'')}(\phi')\bigr)+\bigl(
  \cEv_{\square(\phi')}(\square)(\phi'')
    -\cEv_{\square(\phi'')}(\square)(\phi')\bigr).\]
In the first summand we have used the permutability of
evolutionary derivations and operators in total derivatives. The second
summand hits the image of~$\square$ by construction. 
Therefore the bracket $[\phi',\phi'']_{\square}$ equals
\begin{equation}\label{EqOplusBKoszul}
[\phi',\phi'']_\square=\cEv_{\square(\phi')}(\phi'')
  -\cEv_{\square(\phi'')}(\phi')+
  \{\!\{\phi',\phi''\}\!\}_\square.
\end{equation}
It contains the two standard summands and
the skew\/-\/symmetric bilinear bracket~$\{\!\{\,,\,\}\!\}_\square$.

\begin{example}
A self\/-\/adjoint (hence non\/-\/Hamiltonian) zero\/-\/order
Noether operator~$A_0$ for~\eqref{d-B} was found in~\cite{JK3Bous}:
\begin{equation}\label{A0}
A_0=\begin{pmatrix} ww_x+v_x & -3u_xw-uw_x & u_x\\
-3u_xw-uw_x & -3w^2w_x-4v_xw-uu_x & v_x\\
u_x & v_x & w_x \end{pmatrix}.
\end{equation}
We discover that its image in 
$\sym\cE$ is involutive; the components of the arising bracket $\ib{\vec{p}}{\vec{q}}{A_0}$
with $\vec{p},\vec{q}\in\cosym\cE$ are
\begin{align*}
\ib{\vec{p}}{\vec{q}}{A_0}^u&= p^w_xq^u-p^uq^w_x
 +3w(p^uq^v_x -p^v_xq^u) +3w(p^vq^u_x-p^u_xq^v)\notag\\
{}&\quad{} +p^u_xq^w-p^wq^u_x
  +2w_x(p^uq^v-p^vq^u)+u(p^vq^v_x-p^v_xq^v),\\
\ib{\vec{p}}{\vec{q}}{A_0}^v&= p^u_xq^u-p^uq^u_x +4w(p^vq^v_x-p^v_xq^v)
 +p^v_xq^w-p^wq^v_x +p^w_xq^v-p^vq^w_x,\\
\ib{\vec{p}}{\vec{q}}{A_0}^w&= u(p^uq^v_x-p^v_xq^u) +3w^2(p^vq^v_x-p^v_xq^v)
 +2u_x(p^vq^u-p^uq^v)\notag\\
{}&\quad{}+w(p^u_xq^u-p^uq^u_x)
 +u(p^vq^u_x-p^u_xq^v) +p^w_xq^w-p^wq^w_x.
\end{align*}
\end{example}

\begin{lemma}[\cite{Olver,Opava}]
The image of any Hamiltonian operator~$\smash{\hat{A}}=
\bigl\|\sum\nolimits_\tau A^{\alpha\beta}_\tau\cdot D_\tau\bigr\|$ is closed w.r.t.\ the commutation\textup{;} the bracket $\ib{\,}{\,}{\hat{A}}$ on its domain equals
\begin{equation}\label{EqDogma}
\ib{\phi'}{\phi''}{\hat{A}}^i=\sum_{\sigma,\alpha} (-1)^\sigma
 \Bigl(D_\sigma\circ\Bigl[\sum_{\tau,\beta} D_\tau(\phi'_\beta)\cdot
 \frac{\dd A_\tau^{\alpha\beta}}{\dd u^i_\sigma}\Bigr]\Bigr)
 \bigl(\phi''_\alpha\bigr).
\end{equation}
\end{lemma}

\begin{example}
The pair $\smash{\bigl(\hat{A}_1,\hat{A}_2\bigr)}$ of compatible Hamiltonian operators for~\eqref{d-B} was obtained in~\cite{Nutku}:
\begin{align}
\hat{A}_1&=\begin{pmatrix} D_x & 0 & 0\\ 0 & -4wD_x-2w_x & D_x \\ 0 & D_x & 0
    \end{pmatrix},\label{A1}\\
\hat{A}_2&=\begin{pmatrix}
(2w^2+4v)\,D_x+2(ww_x+v_x) & -11uw\,D_x-(5u_xw+9uw_x) & 3u\,D_x +u_x\\
-11uw\,D_x-6u_xw-2uw_x & 2h\,D_x+h_x & 4v\,D_x +v_x \\
3u\,D_x+2u_x & 4v\,D_x+3v_x & 2w\,D_x +w_x
\end{pmatrix},\label{A2}
\end{align}
where we put 
$h=-(\tfrac{3}{2}u^2+8vw+3w^3)$.
The components of the brackets $\ib{\,}{\,}{\hat{A}_i}$ are given 
by~\eqref{EqDogma}: for any $\vec{p},\vec{q}\in\cosym\cE$ 
they equal, respectively,
\begin{equation}\label{A1Bracket}
\ib{\,}{\,}{\hat{A}_1}^u =\ib{\,}{\,}{\hat{A}_1}^v=0,\qquad
\ib{\vec{p}}{\vec{q}}{\hat{A}_1}^w=2\bigl(p^vq^v_x-p^v_xq^v\bigr)
\end{equation}
and
\begin{subequations}\label{A2Bracket}
\begin{align}
\ib{\vec{p}}{\vec{q}}{\hat{A}_2}^u&= 2(p^w_xq^u-p^uq^w_x)
 +6w(p^uq^v_x -p^v_xq^u) +5w(p^vq^u_x-p^u_xq^v)\notag\\
{}&\quad{} +p^u_xq^w-p^wq^u_x
  +4w_x(p^uq^v-p^vq^u)+3u(p^vq^v_x-p^v_xq^v),
\\
\ib{\vec{p}}{\vec{q}}{\hat{A}_2}^v&= 
 2(p^u_xq^u-p^uq^u_x) +8w(p^vq^v_x-p^v_xq^v)
 +p^v_xq^w-p^wq^v_x +3(p^w_xq^v-p^vq^w_x),
\\
\ib{\vec{p}}{\vec{q}}{\hat{A}_2}^w&= 2u(p^uq^v_x-p^v_xq^u) 
 +(8v+9w^2)\cdot(p^vq^v_x-p^v_xq^v)  +4u_x(p^vq^u-p^uq^v)\notag\\
{}&\quad{}+2w(p^u_xq^u-p^uq^u_x)
 +9u(p^vq^u_x-p^u_xq^v) +p^w_xq^w-p^wq^w_x.
\end{align}
\end{subequations}
We claim that the Noether operator~$A_0$, whose image in $\sym\cE$ is closed w.r.t.\ the com\-mu\-ta\-ti\-on, is compatible with~$\smash{\hat{A}_1}$ 
and~$\smash{\hat{A}_2}$ in this sense.
\end{example}

\begin{defNo}
We say that $N\geq2$ operators $A_i\colon\Cosym\cE\to\sym\cE$ on~$\cE$ with 
a common domain and involutive images, $[\img A_i,\img A_i]\subseteq\img A_i$
for $1\leq i\leq N$, are \emph{linear compatible} 
if their linear combinations $A_{\vec{\lambda}}=\sum_{i=1}^N\lambda_i A_i$
retain the same property of involutivity for any~$\vec{\lambda}$.
\end{defNo}

\begin{example}
It can easily be checked that three Noether operators~\eqref{A0}, 
\eqref{A1}, and~\eqref{A2} for system~\eqref{d-B} are linear compatible.
\end{example}

We note that linear compatible Hamiltonian operators are Poisson compatible,
and \emph{vice versa}, because formula~\eqref{EqDogma} is linear 
in coefficients of~$\smash{\hat{A}}$.

\begin{theor}
The bracket $\ib{\,}{\,}{A_{\vec{\lambda}}}$ on the domain of the 
combination $A_{\vec{\lambda}}$ of linear compatible operators~$A_i$ is
\[
\ib{\,}{\,}{\sum\limits_{i=1}^N\lambda_i A_i}=
 \sum_{i=1}^N\lambda_i\cdot\ib{\,}{\,}{A_i}.
\]
The pairwise linear compatibility implies the collective linear compatibility
of~$A_1,\ldots,A_N$. 
\end{theor}

\begin{proof}
This is readily seen by inspecting the coefficients of $\lambda_i^2$ in
the quadratic polynomials in $\lambda_i$ that appear in both sides of
the equality $\bigl[A_{\vec{\lambda}}(p),A_{\vec{\lambda}}(q)\bigr]=
A_{\vec{\lambda}}\bigl([p,q]_{A_{\vec{\lambda}}}\bigr)$,
here $p,q\in\Omega=\dom A_i/\bigcap_{j=1}^N\ker A_j$
for any~$i$. 

Consider the commutator $\bigl[\sum_i\lambda_iA_i(p),\sum_j\lambda_j A_j(q)\bigr]$.
On one hand, it is equal to
\begin{align}
{}&=\sum_{i\neq j}\lambda_i\lambda_j\bigl[A_i(p),A_j(q)\bigr]+
  \sum_i\lambda_i^2 A_i\bigl(\cEv_{A_i(p)}(q)-\cEv_{A_i(q)}(p)+\ib{p}{q}{A_i}\bigr).
\label{SqLambdaApart}\\
\intertext{On the other hand, the linear compatibility of~$A_i$ implies}
{}&=A_{\vec{\lambda}} 
\bigl(\cEv_{A_{\vec{\lambda}} 
(p)}(q)\bigr)  -  A_{\vec{\lambda}}   
\bigl(\cEv_{A_{\vec{\lambda}}   
(q)}(p)\bigr)  +  A_{\vec{\lambda}}   
\bigl(\ib{p}{q}{A_{\vec{\lambda}}   
}\bigr).\notag
\end{align}
The entire commutator is quadratic homogeneous in~$\vec{\lambda}$, whence the
bracket $\ib{\,}{\,}{A_{\vec{\lambda}}}$ is linear in~$\vec{\lambda}$.
From~\eqref{SqLambdaApart} we see that the individual brackets~$\ib{\,}{\,}{A_i}$
are contained in~it. Therefore,
\[
\ib{p}{q}{A_{\vec{\lambda}}}=\sum_\ell\lambda_\ell\cdot\ib{p}{q}{A_\ell}+
  \sum_\ell\lambda_\ell\cdot\gamma_\ell(p,q),
\]
where $\gamma_\ell\colon\Omega
\times\Omega
\to\Omega
$. We claim that all summands $\gamma_\ell(\cdot,\cdot)$,
which do not depend on $\vec{\lambda}$ at all, vanish. Indeed, assume the converse.
Let there be $\ell\in[1,\ldots,N]$ such that $\gamma_\ell(p,q)\neq0$; without loss
of generality, suppose $\ell=1$. Then set $\vec{\lambda}=(1,0,\ldots,0)$, whence
\begin{multline*}
\Bigl[\sum_i\lambda_iA_i(p),\sum_j\lambda_jA_j(q)\Bigr]=
\Bigl[\bigl(\lambda_1A_1\bigr)(p),\bigl(\lambda_1A_1\bigr)(q)\Bigr]
=\bigl(\lambda_1A_1\bigr)\bigl(\lambda_1\gamma_1(p,q)\bigr)\\
{}+\bigl(\lambda_1A_1\bigr)\Bigl(\cEv_{(\lambda_1A_1)(p)}(q)-
   \cEv_{(\lambda_1A_1)(q)}(p)+ 
    \lambda_1\ib{p}{q}{A_1} 
\Bigr).
\end{multline*}
Consequently, $\gamma_\ell(p,q)\in\ker A_\ell$ for all $p$ and~$q$.
Now we are forced to use the 
nondegeneracy assumption\footnote{We assume that there is no functional freedom
in the kernels of~$A_\ell$. This restriction and its consequences
will be discussed in full detail in a separate paper.} 
$\bigcap_\tau\ker A_\ell^\tau=\{0\}$
for the operators $A_\ell=\|\sum_\tau A_\ell^\tau\cdot D_\tau\|$.
Finally, we have that $\gamma_\ell=0$ for all $\ell$, 
which concludes the proof.
\end{proof}

\begin{cor}
Two such operators~$A$ and~$B$ with involutive images are linear compatible
iff for any $p$,\ $q\in\Omega$ 
one has
\[
\bigl[B(p),A(q)\bigr]+\bigl[A(q),B(p)\bigr]=
 A\bigl([p,q]_B\bigr)+B\bigl([p,q]_A\bigr),
\]
which is equivalent to the relation
\[
\cEv_{A(p)}(B)(q)+\cEv_{B(p)}(A)(q)-\cEv_{A(q)}(B)(p)-\cEv_{B(q)}(A)(p)=
 A\bigl(\ib{p}{q}{B}\bigr)+B\bigl(\ib{p}{q}{A}\bigr).
\]
\end{cor}

\subsection*{Recursion operators and new Lie brackets}
The second step in the construction of the new Lie brackets on~$\sym\cE$
is as follows. We note that the zero\/-\/order 
operator~\eqref{A0} has the inverse $\omega=A_0^{-1}$ 
on an open dense subset of~$\cE$. This yields the local recursion 
operators $R_i\mathrel{{:}{=}}\smash{\hat{A}_i\circ A_0}
\colon\sym\cE\to\sym\cE$.
By construction, their images are closed w.r.t.\ the commutation~$[\,,\,]$. 
(We note also that the images of the Hamiltonian operators~$\smash{\hat{A}_i}$
for~$\cE$ are generally not closed w.r.t.~$[\,,\,]_{R_j}$, 
here $1\leq i,j\leq2$.)
Therefore we determine the new Lie algebra structures $[\,,\,]_{R_i}$ on~$\sym\cE$ by~\eqref{YB}. Let us summarize the result.

\begin{state}
The recursion operators $R_i=\smash{\hat{A}_i\circ A_0^{-1}}$, $i=1,2$, 
for system~\eqref{d-B} are linear compatible. 
The new Lie algebra structures $[\,,\,]_{R_i}$ on~$\sym\cE$ span
the two\/-\/dimensional space of compatible nontrivial finite deformations 
of the standard bracket~$[\,,\,]$.
\end{state}

The transformation rules for $R_i$ and hence for~$[\,,\,]_{R_i}$ under any reparametrizations of the variables $u,v,w,$ are obvious. 
The arising bi\/-\/differential brackets $\ib{\,}{\,}{R_i}$, 
see~\eqref{EqOplusBKoszul}, are obtained from $\ib{\,}{\,}{\hat{A}_i}$ 
using the following `chain rule' (c.f.~\cite{SandersJPW}).

\begin{theor}\label{ChainRule}
Suppose that the images of linear differential operators~$A$
and $R= A\circ\omega$ are closed w.r.t.\ the commutation
of evolutionary vector fields. Then the brackets ${\{\!\{\,,\,\}\!\}}_{A}$
and ${\{\!\{\,,\,\}\!\}}_{R}$ are related by the formula
\begin{equation}\label{ChainRuleFormula}
\omega ({\{\!\{\xi_1,\xi_2\}\!\}}_{R})=
 \cEv_{R(\xi_1)}(\omega)(\xi_2)-\cEv_{R(\xi_2)}(\omega)(\xi_1)
 +{\{\!\{\omega(\xi_1),\omega(\xi_2)\}\!\}}_{A}
\end{equation}
for any sections $\xi_1,\xi_2$ that belong to 
the domain of~$\omega$.
\end{theor}

\begin{proof} 
Denote $\psi_i=\omega(\xi_i)$ and $\vph_i=A(\psi_i)$ for $i=1,2$. We have
\begin{subequations}
\begin{align}
[\varphi_1,\varphi_2]&=
(A\circ\omega)\bigl(\cEv_{\varphi_1}(\xi_2)-\cEv_{\varphi_2}(\xi_1)
 +\ib{\xi_1}{\xi_2}{A\circ\omega}\bigr).\label{Bottom}\\
\intertext{On the other hand, we recall that $\psi_i=\omega(\xi_i)$
and deduce}
[\varphi_1,\varphi_2]&=
A\bigl(\cEv_{\varphi_1}(\psi_2)-\cEv_{\varphi_2}(\psi_1)
 +\ib{\psi_1}{\psi_2}{A}\bigr)\label{Medium}\\
{}&=(A\circ\omega)
  \bigl(\cEv_{\varphi_1}(\xi_2)-\cEv_{\varphi_2}(\xi_1)\bigr)
 +A\bigl(\cEv_{\varphi_1}(\omega)(\xi_2)
   -\cEv_{\varphi_2}(\omega)(\xi_1)+\ib{\psi_1}{\psi_2}{A}\bigr).\notag
\end{align}
\end{subequations}
Now subtract~\eqref{Bottom} from~\eqref{Medium}.\ %
Omitting the operator~$A$, we obtain the assertion.
\end{proof}

The bi\/-\/differential brackets $\ib{\,}{\,}{R_i}$ for the recursions~$R_i$
constructed above are completely determined by the chain 
rule~\eqref{ChainRuleFormula} with~$\omega=A_0^{-1}$:
\begin{equation}\label{BrREnd}
\ib{\vph_1}{\vph_2}{R_i} = A_0\Bigl(
 \cEv_{R_i(\vph_1)}(\omega)(\vph_2)-\cEv_{R_i(\vph_2)}(\omega)(\vph_1)
 +\ib{\omega(\vph_1)}{\omega(\vph_2)}{\hat{A}_i}\Bigr),
\end{equation}
here $\vph_1,\vph_2\in\sym\cE$ are any symmetries of~\eqref{d-B}.
The three components of each bracket $\ib{\,}{\,}{R_i}$ can be
calculated explicitly, e.g., using the environment~\cite{Jets}, 
see Appendix~\ref{AppRecJets}. 
The coefficients of the skew\/-\/symmetric 
couplings~$D_x^\alpha(\vph_1^a)\cdot D_x^\beta(\vph_2^b)-
D_x^\beta(\vph_1^b)\cdot D_x^\alpha(\vph_2^a)$, $0\leq\alpha+\beta\leq1$, 
$1\leq a,b\leq 3$, are relatively big due to the presence of the powers
${(\det A_0)}^\alpha$, $1\leq\alpha\leq 3$ in the denominators. 
Be that as it may, the two local recursion
operators~$R_i$ generate the first known examples
of compatible well\/-\/defined new Lie brackets on the symmetry algebra 
of system~\eqref{d-B} via the Yang\/--\/Baxter equation~\eqref{YB}.

\begin{rem}
The application of the classical $r$-\/matrix formalism~\cite{ReymanSTSh} 
for a given Lie algebra~$\gothg$
generates Liouville integrable systems using the second Lie algebra structure $[\,,\,]_{r}$ that solves the Yang\/--\/Baxter equation~YB($\alpha$).
Since equation~\eqref{YB} corresponds to the degenerate case $\alpha=0$, 
we pose the problem of finding recursion operators for $\gothg=\sym\cE$
that will lead to relevant factorizations and produce new integrable systems. 
\end{rem}

\begin{rem}
Particular examples of non\/-\/Hamiltonian linear operators with involutive images, and Lie brackets on their domains, are scattered in the literature
(e.g., see~\cite{Bogoyavl} for a dispersionless set\/-\/up and a bracket of $1$-\/forms). However, let us remember that, first, the differential order of
such operators can be sufficiently high for systems
with dispersion. Second, the domains and images of such operators can be composed by (co)symmetries of two \emph{different} equations. For example, a class of higher\/-\/order operators with involutive images is known for the
open 2D~Toda chains and the related KdV\/-\/type systems, 
see~\cite{TMPhGallipoli}. 
Involutive distributions of operator\/-\/valued evo\-lu\-ti\-o\-na\-ry
vector fields will be the object of a subsequent publication.
\end{rem}

\subsection*{Acknowledgements}
The authors thank 
I.\,S.\,Kra\-sil'\-sh\-chik and V.\,V.\,Sokolov
for useful discussions. 
This work has been partially supported by the European Union through
the FP6 Marie Curie RTN \emph{ENIGMA} (Contract
no.\,MRTN-CT-2004-5652), the European Science Foundation Program
{MISGAM}, and by NWO grants~B61--609 and VENI~639.031.623.
A.\,K.\ thanks $\smash{\text{IH\'ES}}$ for financial support 
and warm hospitality.

\appendix

\section{Calculation of $\ib{\,}{\,}{R_i}$ for the two
recursions~$R_i$}\label{AppRecJets}\noindent%
The following program for the \textsc{Jets} environment~\cite{Jets}
under \textsc{Maple} calculates the bracket $\ib{\,}{\,}{R_1}$
on the domain of the recursion 
operator~$R_1=\smash{\hat{A}_1\circ A_0^{-1}}$, see~\eqref{A0} and~\eqref{A1},
for the dispersionless $3$-\/component Boussinesq\/-\/type system~\eqref{d-B}.
The bracket $\ib{\,}{\,}{R_2}$ induced by~$R_2$ is obtained using a
slight modification of this program. The modification amounts to 
a substitution of~$\smash{\hat{A}_2}$ and $\ib{\,}{\,}{\hat{A}_2}$, 
see~\eqref{A2Bracket}, for $\smash{\hat{A}_1}$ and $\ib{\,}{\,}{\hat{A}_1}$,
respectively.
\begin{verbatim}
> read `Jets.s`;
> coordinates([x],[u,v,w,p1,p2,p3,q1,q2,q3],5):
\end{verbatim}
The dependent coordinates $\mathtt{p}^i$, $\mathtt{q}^j$ denote
the components of the sections 
$\vph_1,\vph_2$ in the domain~$\sym\cE$ of~$R_1$.
The notation is in parallel with formula~\eqref{BrREnd}.
\begin{verbatim}
> A:=Matrix(3,3): ia:=Matrix(3,3): M:=Matrix(3,3): N:=Matrix(3,3):
> A:=<<w*w_x+v_x,-3*w*u_x-u*w_x,u_x>|
  <-3*w*u_x-u*w_x,-3*w^2*w_x-4*w*v_x-u*u_x,v_x>|<u_x,v_x,w_x>>;
\end{verbatim}
We have assigned $\mathtt{A}=A_0$, see~\eqref{A0}.
\begin{verbatim}
> with(LinearAlgebra):
> ia:=MatrixInverse(A):
\end{verbatim}
This yields $\mathtt{ia}=A_0^{-1}\colon\sym\cE\to\cosym\cE$, which we also denote
by~$\omega$. Now put $\mathtt{m}=A_0^{-1}(\vph_1)$ and $\mathtt{n}=A_0^{-1}(\vph_2)$.
\begin{verbatim}
> m1 := simplify(ia[1,1]*p1+ia[1,2]*p2+ia[1,3]*p3):
> m2 := simplify(ia[2,1]*p1+ia[2,2]*p2+ia[2,3]*p3):
> m3 := simplify(ia[3,1]*p1+ia[3,2]*p2+ia[3,3]*p3):
>
> n1 := simplify(ia[1,1]*q1+ia[1,2]*q2+ia[1,3]*q3):
> n2 := simplify(ia[2,1]*q1+ia[2,2]*q2+ia[2,3]*q3):
> n3 := simplify(ia[3,1]*q1+ia[3,2]*q2+ia[3,3]*q3):
\end{verbatim}
We have computed $\psi_i=\omega(\vph_i)$; we recall that $\psi\in\cosym\cE$.

Next, we apply the Hamiltonian operator~$\smash{\hat{A}_1}$ and obtain
symmetries of~\eqref{d-B},
$\mathtt{k}=\smash{\hat{A}_1}(\mathtt{m})=\smash{\hat{A}_1}(A_0^{-1}(\vph_1))$
and $\mathtt{l}=\smash{\hat{A}_1}(\mathtt{n})=\smash{\hat{A}_1}(A_0^{-1}(\vph_2))$.
\begin{verbatim}
> k1 := simplify(evalTD(TD(m1,x))):
> k2 := simplify(evalTD(-4*w*TD(m2,x)-2*w_x*m2+TD(m3,x))):
> k3 := simplify(evalTD(TD(m2,x))):
>
> l1:=simplify(evalTD(TD(n1,x))):
> l2:=simplify(evalTD(-4*w*TD(n2,x)-2*w_x*n2+TD(n3,x))):
> l3:=simplify(evalTD(TD(n2,x))):
\end{verbatim}

Now we act by evolutionary derivations on the coefficients of
the operator~$\omega=A_0^{-1}$, see~\eqref{ChainRuleFormula} 
and~\eqref{BrREnd}. We set $\mathtt{M}=\cEv_{R_1(\vph_1)}(\omega)$; 
note that the matrix~$\mathtt{M}$ is symmetric.
\begin{verbatim}
> M[1,1]:=simplify(evalTD(k1*pd(ia[1,1],u)+TD(k1,x)*pd(ia[1,1],u_x)+
  k2*pd(ia[1,1],v)+TD(k2,x)*pd(ia[1,1],v_x)+k3*pd(ia[1,1],w)+
  TD(k3,x)*pd(ia[1,1],w_x))):
> M[1,2]:=simplify(evalTD(k1*pd(ia[1,2],u)+TD(k1,x)*pd(ia[1,2],u_x)+
  k2*pd(ia[1,2],v)+TD(k2,x)*pd(ia[1,2],v_x)+k3*pd(ia[1,2],w)+
  TD(k3,x)*pd(ia[1,2],w_x))):
> M[1,3]:=simplify(evalTD(k1*pd(ia[1,3],u)+TD(k1,x)*pd(ia[1,3],u_x)+
  k2*pd(ia[1,3],v)+TD(k2,x)*pd(ia[1,3],v_x)+k3*pd(ia[1,3],w)+
  TD(k3,x)*pd(ia[1,3],w_x))):
> M[2,2]:=simplify(evalTD(k1*pd(ia[2,2],u)+TD(k1,x)*pd(ia[2,2],u_x)+
  k2*pd(ia[2,2],v)+TD(k2,x)*pd(ia[2,2],v_x)+k3*pd(ia[2,2],w)+
  TD(k3,x)*pd(ia[2,2],w_x))):
> M[2,3]:=simplify(evalTD(k1*pd(ia[2,3],u)+TD(k1,x)*pd(ia[2,3],u_x)+
  k2*pd(ia[2,3],v)+TD(k2,x)*pd(ia[2,3],v_x)+k3*pd(ia[2,3],w)+
  TD(k3,x)*pd(ia[2,3],w_x))):
> M[3,3]:=simplify(evalTD(k1*pd(ia[3,3],u)+TD(k1,x)*pd(ia[3,3],u_x)+
  k2*pd(ia[3,3],v)+TD(k2,x)*pd(ia[3,3],v_x)+k3*pd(ia[3,3],w)+
  TD(k3,x)*pd(ia[3,3],w_x))):
\end{verbatim}
In the same way, we define the symmetric matrix
$\mathtt{N}=\cEv_{R_1(\vph_2)}(\omega)$.
\begin{verbatim}
> N[1,1]:=simplify(evalTD(l1*pd(ia[1,1],u)+TD(l1,x)*pd(ia[1,1],u_x)+
  l2*pd(ia[1,1],v)+TD(l2,x)*pd(ia[1,1],v_x)+l3*pd(ia[1,1],w)+
  TD(l3,x)*pd(ia[1,1],w_x))):
> N[1,2]:=simplify(evalTD(l1*pd(ia[1,2],u)+TD(l1,x)*pd(ia[1,2],u_x)+
  l2*pd(ia[1,2],v)+TD(l2,x)*pd(ia[1,2],v_x)+l3*pd(ia[1,2],w)+
  TD(l3,x)*pd(ia[1,2],w_x))):
> N[1,3]:=simplify(evalTD(l1*pd(ia[1,3],u)+TD(l1,x)*pd(ia[1,3],u_x)+
  l2*pd(ia[1,3],v)+TD(l2,x)*pd(ia[1,3],v_x)+l3*pd(ia[1,3],w)+
  TD(l3,x)*pd(ia[1,3],w_x))):
> N[2,2]:=simplify(evalTD(l1*pd(ia[2,2],u)+TD(l1,x)*pd(ia[2,2],u_x)+
  l2*pd(ia[2,2],v)+TD(l2,x)*pd(ia[2,2],v_x)+l3*pd(ia[2,2],w)+
  TD(l3,x)*pd(ia[2,2],w_x))):
> N[2,3]:=simplify(evalTD(l1*pd(ia[2,3],u)+TD(l1,x)*pd(ia[2,3],u_x)+
  l2*pd(ia[2,3],v)+TD(l2,x)*pd(ia[2,3],v_x)+l3*pd(ia[2,3],w)+
  TD(l3,x)*pd(ia[2,3],w_x))):
> N[3,3]:=simplify(evalTD(l1*pd(ia[3,3],u)+TD(l1,x)*pd(ia[3,3],u_x)+
  l2*pd(ia[3,3],v)+TD(l2,x)*pd(ia[3,3],v_x)+l3*pd(ia[3,3],w)+
  TD(l3,x)*pd(ia[3,3],w_x))):
\end{verbatim}
We act by the operators $\mathtt{M},\mathtt{N}$ on $\vph_2$ and $\vph_1$,
respectively, and calculate the difference
$\mathtt{e}=\cEv_{R_1(\vph_1)}(\omega)(\vph_2)
  -\cEv_{R_1(\vph_2)}(\omega)(\vph_1)$.
\begin{verbatim}
> e1:=simplify(M[1,1]*q1+M[1,2]*q2+M[1,3]*q3-N[1,1]*p1-N[1,2]*p2-N[1,3]*p3):
> e2:=simplify(M[1,2]*q1+M[2,2]*q2+M[2,3]*q3-N[1,2]*p1-N[2,2]*p2-N[2,3]*p3):
> e3:=simplify(M[1,3]*q1+M[2,3]*q2+M[3,3]*q3-N[1,3]*p1-N[2,3]*p2-N[3,3]*p3):
\end{verbatim}

Next, we recall that the bracket $\ib{\,}{\,}{\hat{A}_1}$ for~$\smash{\hat{A}_1}$
is equal to~\eqref{A1Bracket}, and substitute $\psi_i=\omega(\vph_i)$ in it.
Thus we put
$\mathtt{s}=\mathtt{e}+\{\!\{\mathtt{m},\mathtt{n}\}\!\}_{\hat{A}_1}$.
\begin{verbatim}
> s1:=simplify(e1):
> s2:=simplify(e2):
> s3:=simplify(e3+evalTD(2*(m2*TD(n2,x)-TD(m2,x)*n2))):
\end{verbatim}
We can now check that formulas~\eqref{A1Bracket} is correct.
\begin{verbatim}
> J:=Jacobi([k1,k2,k3,0,0,0,0,0,0],[l1,l2,l3,0,0,0,0,0,0]):
> D1:=simplify(evalTD(TD(s1,x))):
> D2:=simplify(evalTD(-4*w*TD(s2,x)-2*w_x*s2+TD(s3,x))):
> D3:=simplify(evalTD(TD(s2,x))):
> simplify(evalTD(J[1]-D1));
                                  0
> simplify(evalTD(J[2]-D2));
                                  0
> simplify(evalTD(J[3]-D3));
                                  0
\end{verbatim}

Finally, we act onto the right\/-\/hand side of~\eqref{ChainRuleFormula}
by the operator~$\omega^{-1}=A_0$, and thus we obtain the components
$\mathtt{Z}=A_0(\mathtt{s})$ of the bracket~$\ib{\,}{\,}{R_1}$.
\begin{verbatim}
> Z1:=simplify((w*w_x+v_x)*s1+(-3*w*u_x-u*w_x)*s2+u_x*s3);
> Z2:=simplify((-3*w*u_x-u*w_x)*s1+(-3*w*w*w_x-4*w*v_x-u*u_x)*s2+v_x*s3);
> Z3:=simplify(u_x*s1+v_x*s2+w_x*s3);
\end{verbatim}
The result is somewhat a surprise: the output contains more than 15,000
lines. 
However, decomposition~\eqref{BrREnd} makes the expressions manageable.
Obviously, the number of essential summands is reduced twice by the skew\/-\/symmetry. Next, the summands in the three components of $\ib{\,}{\,}{R_1}$ 
are collected at the nine bilinear expressions
$p^i\cdot q_x^j-p_x^j\cdot q^i$, $1\leq i,j\leq3$ 
and at the six multiples $p^i\cdot q^j - p^j\cdot q^i$ 
with $1\leq i\neq j\leq3$. All the coefficients
are fractions of differential polynomials in $u,v,w$ with 
$(\det A_0)^3$ in the denominators; many of these fractions are reducible.

\end{document}